\documentstyle[aps]{revtex}

\begin{document}

\title{Exchange and correlation energies of ground states of
atoms and molecules in strong magnetic fields
}
\author{P. Schmelcher$^{1}$, M. V. Ivanov$^{2}$ and W. Becken$^{1}$ 
}
\address{
$^{1}$
Theoretische Chemie, Physikalisch--Chemisches Institut,
Universit\"at Heidelberg, INF 253, D-69120 Heidelberg,
Federal Republic of Germany\\
}
\address{
$^{2}$
Institute of Precambrian Geology and Geochronology,
Russian Academy of Sciences,
Nab. Makarova 2, St. Petersburg 199034, Russia\\
}
\maketitle

\begin{abstract}
Using a Hartree-Fock mesh method and a configuration interaction approach
based on a generalized Gaussian basis set we investigate the behaviour
of the exchange and correlation energies of small atoms and molecules, namely the helium and
lithium atom as well as the hydrogen molecule, in the presence of a 
magnetic field covering the regime $B=0-100a.u.$   
In general the importance of the exchange energy to the binding properties of atoms
or molecules increases strongly with increasing field strength.
This is due to the spin-flip transitions
and in particular due to the contributions of the tightly bound hydrogenic states
which are involved in the corresponding ground states of different symmetries.
In contrast to the exchange energy the correlation energy becomes less
relevant with increasing field strength. This holds for the individual
configurations constituting the ground state and for the crossovers of
the global ground state.

\end{abstract}


\section{Introduction}

Particle systems in strong magnetic fields represent a source of
complex nonlinear phenomena which have their origin in the competing
particle-particle and magnetic interactions. Considering the 
simplest of all possible systems, namely the hydrogen atom \cite{Fri89}, we
encounter a variety of different properties and qualitatively
new effects. Leaving aside the inherent two-body character of the system
\cite{Schm96} let us concentrate on the evolution of the electronic properties 
with increasing field strength. Since the symmetry gradually changes with
increasing field strength from a field-free spherical to a cylindrical one
in the high field limit we encounter a strong rearrangement of the
electronic wave function which is particularly dramatic in the so-called intermediate
regime for which the strength of the two interactions are of comparable
order of magnitude. The large number of investigations on the hydrogen
atom in a magnetic field during the past twenty years provided us with a
detailed understanding of the quantum properties of the atom.
It thereby became a prominent example for a low-dimensional chaotic dynamical system
as well as a testing ground for the development of semiclassical theories
of nonintegrable systems \cite{Mai97}.

In contrast to the hydrogen atom our knowledge on the behaviour and properties of
many electron atoms in strong magnetic fields is still in its infancy.
There exists a moderate number of investigations
on the ground and low-lying excited states of the helium atom for field strengths
ranging from the weak to the high field regime (see refs.\cite{Rud94,Jon97,Bra98,Iva94}
and references therein). Very recently precise calculations of a considerable number
of excited states of different symmetries \cite{Bec98} were performed for a wide range of
astrophysically relevant field strengths. As a result a successful comparison of the 
corresponding stationary transitions with the astronomically observed spectrum of the
magnetic white dwarf GD229 could be performed resulting in a strong evidence for
the existence of helium in the atmosphere of this object \cite{Jor98}.
For atoms with more than two electrons there exist only few investigatons
of partially qualitative character which focus almost exclusively on the high field
regime (see, for example, ref.\cite{Mil91} and references therein).
For the lithium atom there are two recent investigations \cite{Jon96,Iva98} which cover also
the intermediate regime of field strengths.

In the case of molecules the situation is somewhat similar. The one-electron problem,
i.e. the hydrogen molecular ion $H_2^+$ in a magnetic field,
has been studied in some detail (see refs.\cite{Kap95} and references therein)
including the situation of an arbitrary orientation of the internuclear axis with respect
to the magnetic field axis. A careful quantitative investigation of the hydrogen molecule
including a correct description of the ground state for arbitrary field strengths,
has however been performed only very recently (see refs.\cite{Kra97,Det97,Det98}).
Nevertheless a number of interesting features have been detected like, for example,
magnetic field induced chemical binding mechanisms or crossovers of ground states 
due to spin or internal energy shifts. 

In view of the above-described dramatic changes of the electronic structure
and properties of atoms and molecules in the presence of a magnetic field it is natural
to ask the question how the different contributions to the total energies change with
changing field strength. More precisely: how does the exchange and correlation
energy behave if the field is turned on and increased from the weak field 
to the high field regime ? This question is of immediate relevance to a
better understanding of the binding properties of inhomogeneous electronic systems in strong
fields. Furthermore our investigation
is also motivated by the persisting need of density functionals in the
presence of strong magnetic fields \cite{Vig88,Sku93,Hol97}
whose construction requires detailed information
about the influence of the external field on the electronic structure.
The aim of the present paper is to study the behaviour of the exchange
and correlation energies for a broad range of field strengths
for the ground states of some selected small
atoms and molecules, namely the helium and lithium atoms and the hydrogen molecule. 

In detail we proceed as follows. Section 2 contains a brief discussion of relevant
theoretical aspects as well as an outline of the different computational methods
used to investigate the electronic structure of atoms and molecules in strong
magnetic fields. Since both the theoretical background as well as the
underlying computational techniques are rather sophisticated we focus here only
on some major points which are of immediate relevance to the present investigation.
Section 3 contains a discussion of the results, i.e. the magnetic field dependent
behaviour of the exchange and correlation energies for the above-mentioned 
atomic and molecular two and three-electron systems. Section 4 provides a
summary and the conclusions.

\section{Theoretical aspects and computational methods}

Atoms and molecules in external magnetic fields show a number of intriguing
new phenomena which are not present in field-free space and which manifest themselves
already on the level of the fundamental equations of motion. Examples are
the inseparability of the center of mass and internal motion \cite{Schm96} and the
screened Born-Oppenheimer approach (see ref.\cite{Schm94} and references therein)
which has to be followed in order to ensure the validity of an adiabatic
approximation for molecules in magnetic fields. In the present investigation
which focuses on the ground states of the corresponding systems 
in the field strength regime $0<B<100a.u.$ ($1a.u.$ corresponds to $2.35\cdot
10^{5}Tesla$) it can be safely assumed that the fixed-nucleus approach, i.e. the assumption
of an infinitely heavy nucleus, is a very good zeroth order approximation
to a more sophisticated electronic Hamiltonian. Therefore we arrive at the following general
appearance of the fixed-nucleus Hamiltonian for the symmetric gauge using
atomic units ($m=|e|=\hbar=1$)
\begin{equation}
{\cal H} = \sum_{i=1}^{N} \left(\frac{{\bf p}_i^2}{2}
+ \frac{1}{2}BL_{z_i} + \frac{1}{8} \left({\bf B} \times {\bf r}_i\right)^2
\right) + V
\end{equation}
where we have assumed that the magnetic field is oriented along the $z-$axis.
${\bf r}_i,{\bf p}_i,L_{z_i}$ are the position, canonical momentum and canonical
angular momentum component parallel to the field of the $i-th$ electron, respectively.
$V$ contains all the Coulomb interaction terms, i.e. the electron-nucleus attraction
as well as the electron-electron and nucleus-nucleus (in the case of molecules)
repulsion. In the case of an atom the symmetries of the above Hamiltonian are:
rotations around the magnetic field axis, inversion of all coordinates
with respect to the origin (position of the nucleus). Consequently the total
angular momentum component $\sum_i L_{z_i}$ parallel to the magnetic field is
conserved (with the magnetic quantum number $M$)
and the parity and $z-$parity operator (with eigenvalue $(-1)^{\Pi_z}$)
commute with $\cal H$. Including the
spin multiplicity $2S+1$ we therefore arrive at the spectroscopic notation
$\nu ^{2S+1}M^{(-1)^{\Pi_z}}$
of the electronic states in the presence of a magnetic field where
$\nu$ designates the degree of excitation within a subspace of certain
symmetries. In the case of diatomic molecules the electronic energies
depend in general on both the internuclear distance as well as the angle
$\Theta$ between the internuclear axis and the magnetic field axis
and the symmetries of the electronic Hamiltonian depend strongly on the mutual
orientation of these axes \cite{Schm90}.
Within the present investigation we restrict ourselves to the parallel
configuration ($\Theta = 0^{\circ}$) of the hydrogen molecule which has been shown to be of 
particular importance for the ground state of the diatomic system \cite{Schm90,Det98}.
Choosing as the origin of the internal coordinate system the 
midpoint between the positions of the identical nuclei we again have the parity, $z-$parity and
rotations around the magnetic field axis as symmetries and 
use correspondingly the molecular spectroscopic notation $\nu^{2S+1}M_{u/g}$
where $u/g$ denotes the ungerade-gerade parity, respectively,
and $M=0,1,2,...$ equals the series $\Sigma,\Pi,\Delta,...$.

Solving the Schr\"odinger equation belonging to the Hamiltonian (1) in particular in
the intermediate regime of field strengths poses a hard methodological and
computational problem. This is due to the fact that the diamagnetic and 
Coulomb interaction terms are of qualitatively different nature and yield already
on the one particle level (one-electron atom or molecule) a nonseparable 
problem. Therefore the well-established methods 
for ab initio calculations of atoms and molecules cannot be directly applied
to electronic structure calculations in a strong magnetic field. Within the 
present investigation we use two methods developed during the past years
especially for the situation of strong fields. They have been 
proven to be very accurate and efficient for studies of electronic systems
in external fields in general. These are first an unrestricted Hartree-Fock
(HF) approach based on a numerical mesh method and second a
configuration interaction approach based on a generalized anisotropic
Gaussian basis set. Combining the results of the corresponding calculations we will 
be able to extract the desired information on the behaviour of the exchange
and correlation energies for small atomic and molecular systems with changing
field strength. In the following we provide some key
ideas and features of the above-mentioned two methods. For further details
and technical aspects we refer the reader to the corresponding literature
given below.

Our unrestricted HF approach is formulated 
in cylindrical coordinates $(\rho,\phi,z)$. We assign
to each electron a certain value of the magnetic quantum
number $m_\mu$ and each single-electron wave function
$\Psi_\mu$ depends on the variables $\phi$ and $(\rho,z)$
according to
\begin{eqnarray}
\Psi_\mu(\rho,\phi,z)=(2\pi)^{-1/2}e^{-i m_\mu\phi}\psi_\mu(z,\rho)
\end{eqnarray}
where $\mu$ indicates the numbering of the electrons.
The resulting partial differential equations for $\psi_\mu(z,\rho)$
and the formulae for the Coulomb and exchange potentials
are given in ref.\cite{Iva94}. The one-particle equations
for the wave functions $\psi_\mu(z,\rho)$ are solved
by means of the numerical mesh method described in refs.
\cite{Iva88,Iva94}. The feature which distinguishes the
present calculations from those described in ref.\cite{Iva94}
is the computational technique for the evaluation of the Coulomb
and exchange integrals. Here as well as in
ref.\cite{Iva97} we obtain these potentials as solutions
of the corresponding Poisson equations.
The problem of the boundary conditions for the Poisson equation
as well as the problem of simultaneously solving
Poissons equations and Schr\"odinger-like equations for the wave
functions $\psi_\mu(z,\rho)$ on the same meshes have been discussed
in ref.\cite{Iva94}. We solve these problems
by using special forms of non-uniform meshes.
Solutions to the Poisson equation on separate meshes contain
some errors $\delta_P$ associated with an inaccurate description of the
potential far from the nucleus. However due to the special form
of the function $\delta_P(h)$ for these meshes
(where $h$ is a formal mesh step)
the errors do not show up in the final results for the energy
and other physical quantities, which we obtain by means of the Richardson
extrapolation procedure (polynomial extrapolation to $h=0$
\cite{Iva88,Iva86}).
An additional improvement with respect to the precision of our numerical
calculations of the integrals is achieved by solving the Poisson equation not for
the whole charge distribution but for the total distribution minus
some properly chosen charge distribution
with known analytical solution to the Poisson equation.

Our mesh approach is flexible enough to yield precise
results for arbitrary field strengths. Some minor decrease of the precision
appears in very strong magnetic fields. This phenomenon is due to a growing difference in the
binding energies ${\epsilon_B}_\mu$ of single-electron wave functions belonging to the same
electronic configuration
\begin{eqnarray}
{\epsilon_B}_\mu=(m_\mu+|m_\mu|+2s_{z\mu}+1)\gamma/2-\epsilon_\mu
\end{eqnarray}
where $\epsilon_\mu$ is the single-electron energy and $s_{z\mu}$ is the spin $z$-projection.
This results in big differences with respect to the spatial extension of the density
distribution for different electrons. The precision of our HF results
depends, of course, on the number of mesh nodes and can be improved in calculations
with denser meshes. The absolute accuracy of the total energies, i.e.
the error within our HF mesh approach of the total energies, is
typically of the order of magnitude of $10^{-6}$, i.e they are very well converged.

Our second method for electronic structure calculations in strong magnetic fields
is a configuration interaction (CI) approach based on an atomic basis set
of generalized anisotropic Gaussian orbitals (GAGO). It has originally been
developed and implemented
for molecules \cite{Schm88,Det97,Det98}
and very recently has been specialized and optimized in order to perform
electronic structure calculations of atomic systems \cite{Bec98}.
In contrast to the above HF-approach this method is exact in the sense
that it does not miss the correlation energy: increasing the number of basis functions
allows, in principle, for an arbitrary accurate description of the electronic wave
functions. Together with the above HF-energies this enables us to 
extract the behaviour of the correlation energy as a function of the field strength.

In the remaining part of this section we briefly describe the essential strategy of 
our GAGO based CI approach. For the spatial part of the electronic eigenfunctions of the 
Hamiltonian (1) we use the linear combination of atomic orbitals ansatz,
i.e. we decompose the wave function into orbital configurations which respect the
corresponding symmetries and the Pauli principle. In the case of the 
hydrogen molecule $H_2^+$ molecular orbital are formed as an intermediate step.
The $H_2^+$ molecular orbitals are built
from atomic orbitals centered at each nucleus. As already mentioned a key ingredient
of this procedure is the GAGO basis set. The individual orbitals have to be optimized
in the presence of a magnetic field in order to ensure an efficient and accurate
description of the atomic or molecular properties. The nonlinear optimization of the corresponding
variational exponential parameters has to be performed for each state of a given symmetry and
for each field strength separately and represents therefore a very extensive work.
Finally in order to determine the atomic or molecular electronic wave functions
we use the variational principle which means we solve the resulting generalized
eigenvalue problem via standard methods. It is very well-known that describing the
correlation behaviour accurately is a very hard task to accomplish. Within the
present approach the relative accuracy of the total energies is estimated to be between
$10^{-3}$ and $10^{-5}$ and the errors of the determined correlation
energies (using the corresponding HF results) are at most a few percent. 

\section{Results and discussion}

In the present section we will first focus on the behaviour of the exchange energies
and subsequently correlation energies as a function of the magnetic field
strength for atoms and thereafter for molecules.
We will provide the absolute values of these energies for a grid of values
of the field strength in the regime $B=0-100$ and furthermore, as an 
indicator of the relevance of these energies to the binding energies,
we will show the ratios of these energies and the total
binding energies as a function of the field strength. 

Let us begin with the neutral two-electron atomic system, i.e. the helium atom.
Covering the regime of field strengths $B=0-100$ the helium atom possesses 
ground states of different symmetries. For the regime of field strengths $0<B^{<}_{\sim}0.7$
the ground state is the $1^10^+$ state which evolves continuously from the 
ground state in the absence of a magnetic field.
At $B\approx 0.7$ there occurs a crossover and the new ground state is of 
$1^3(-1)^+$ symmetry which remains the ground state for the complementary regime $B>0.7$,
i.e. for arbitrary high magnetic field strengths. Since the spin singlet $1^10^+$
state possesses no exchange energy we are exclusively interested in the triplet
state $1^3(-1)^+$. Table I gives the absolute values of the exchange energy $E_{ex}$ of
the $1^3(-1)^+$ state for twenty values of the field strength in the regime $B=0-100$.
We observe a monotonous increase from $E_{ex}=0.00644$ for $B=0$ to
$E_{ex}=0.43872$ for $B=100$ i.e. by approximately two orders of magnitude.
This clearly demonstrates there is a strong increase in the absolute value of the
exchange energy with increasing field strength. Since however the binding energy
of the helium atom in the $1^3(-1)^+$ state raises also with increasing field
strength we have to consider the ratio $R_{xb}$ of the exchange energy and the total
binding energy in order to obtain an indicator for the influence of the exchange
energy on the binding properties of the atom with changing field strength.
This is illustrated in Figure 1 where the behaviour
of $R_{xb}$ for the $1^3(-1)^+$ state of the helium atom is shown for $10^{-3}<B<10^{2}$
on a double logarithmic scale. $R_{xb}$ increases by more than one order
of magnitude showing that the importance of the exchange energy for the
binding properties of the $1^3(-1)^+$ state of the atom increases with increasing field strength.

What is the origin of this dramatic change in the exchange energy with increasing 
field strength? To elucidate this we have to take a step back and consider the
hydrogen atom in a magnetic field. It is well-known that each of the lowest 
electronic states of gerade z-parity and a given (negative) magnetic quantum number
shows a monotonous increase in binding energy with increasing field strength
\cite{Avr79}. These states uniquely correspond to the series of field-free hydrogen states
$1s_{0},2p_{-1},3d_{-2},...$. Indeed their binding energy diverges like $E_B \rightarrow
\infty$ in the mathematical limit $B\rightarrow \infty$. At the same time
the expectation value of the radius for these orbitals goes to zero, i.e.
$<r> \rightarrow 0$.
Moreover, the binding energies of all these orbitals have small
differences in the limit $B\rightarrow \infty$. As a result their
expectation values $<r>$ are also close together and this means
that the electron densities corresponding to these orbitals are
concentrated in the same spatial region. On the other hand the binding
energies and the spatial electron density distributions for these states
are very different for $B=0$.
The $1^3(-1)^+$ state of helium consists, in the language of HF,
of the $1s2p_{-1}$ orbitals constituting, according to the above, two strongly bound orbitals
which shrink monotonically with increasing field strength.
Comparing for different field strengths the relative changes of $<r>$ for the two orbitals
$1s$ and $2p_{-1}$ it is obvious that they are much larger for the $2p_{-1}$ orbital. 
In the limit $B\rightarrow \infty$ this orbital overlaps strongly with
the $1s$ orbital.
This causes a strong increase in the overlap of the orbitals involved in the
integral expression for the exchange energy and thereby explains the above-observed
increase of the exchange energy with increasing field strength.

Next let us consider the neutral three-electron atom, i.e. the lithium atom in a 
magnetic field \cite{Iva98}. The field-free ground state possesses $^20^+$ symmetry and, in a 
HF language, is represented by the spin dublet configuration $1s^22s$.
The binding energy of this state increases monotonically with increasing field strength. 
A careful investigation shows\cite{Iva98} that there exist two crossovers with 
respect to the ground state of the Li atom with increasing field strength finally
ending up in a spin quartett state. The first crossover occurs at $B_{c1}=0.176$
up to which the $1^20^+$ remains the global ground state. The new ground state above
$B_{c1}$ is constituted by the $1^2(-1)^+$ state which, in the HF picture, is represented
by the spin dublet configuration $1s^22p_{-1}$. This first change in the ground
state is therefore not associated with a spin flip process but a pure change in the
spatial orbitals: the weakly bound $2s$ orbital is replaced by the strongly bound $2p_{-1}$
orbital. The $1^2(-1)^+$ spin dublet state is the global ground state of the Li atom
in the regime $0.176<B<2.153$. At $B_{c2}=2.153$ a second crossover takes place
and for $B>B_{c2}$ the new ground state is represented by the fully spin polarized
$1^4(-3)^+$ state which is constituted by the $1s2p_{-1}3d_{-2}$ orbital configuration.

Table 1 provides the exchange energies of the
above-mentioned ground states $1^20^+$, $1^2(-1)^+$ and $1^4(-3)^+$ of the lithium
atom, respectively, for the same grid of field strengths used for the helium atom.
Starting with a vanishing field strength the exchange energy for the 'weak-field
ground state' $1^20^+$ first increases with increasing field strength by approximately
a factor two up to $B \approx 1$ and subsequently decreases finally ($B=100$) being lower than the
original field-free value. A similar behaviour can be observed for the corresponding
quantity $R_{xb}(B)$ the drop after the increase being however even larger since
the total binding energy increases with increasing field strength. 
The exchange energies of the $1^2(-1)^+$ and $1^4(-3)^+$ states of the lithium
atom increase monotonically with increasing field strength in the complete
regime $0<B<100$. The factor of increase is approximately
$40$ for the $1^4(-3)^+$ state and even larger, i.e.
$200$, for the $1^2(-1)^+$ state. Analogously the quantity $R_{xb}$ increases
with increasing field strength for both states (see figure 1). The latter 
clearly indicates the increasing importance of the exchange energy for
the binding properties of the atom with increasing field strength.
The explanation provided above for the increase of the exchange energy 
of the helium atom in a magnetic field holds also for the $1^2(-1)^+$ and
$1^4(-3)^+$ states of lithium. The HF-orbitals responsible for the increase 
of the exchange energy are the $1s,2p$ and $1s,2p,3d$ orbitals, respectively.
Obviously this argument does not hold for the $1^20^+$ state since the involved
$2s$ HF-orbital shows no unrestricted shrinking in size with increasing
field strength.

We now investigate the correlation energies for the ground states $1^10^+$ and
$1^3(-1)^+$ of the helium atom. (Unfortunately there exists no approach currently
which would allow to investigate the correlation energy of more than two-electron
systems in the presence of a strong magnetic field).
The corresponding values are given in table 2 for a grid of
eleven field strengths in the regime $0<B<100$.
Let us first consider the 'weak-field ground state' $1^10^+$. Its correlation energy stays
approximately constant up to a field strength of $2$ and then increases only
by a factor of two while increasing the field strength up to $100$.
Figure 2 shows the ratio $R_{cb}$ of the correlation energy and the corresponding
total binding energy. $R_{cb}$ stays approximately constant up to $B=0.1$ and 
decreases thereafter which indicates that the correlation energy becomes 
increasingly unimportant for the binding properties
of the $1^10^+$ state with increasing field strength.
If we focus on the correlation energies of the $1^3(-1)^+$ state
of the helium atom we observe (see table 2) a monotonous increase of roughly one
order of magnitude for the whole regime $B=0-100$. $R_{cb}$ for the $1^3(-1)^+$ state
first increases (see figure 2) with increasing field strength up to $B\approx1$,
then shows a minor decrease and finally increases again for $B^{>}_{\sim}10$.
The absolute values of the correlation energies of this state are, for any field strength, much
smaller than the corresponding exchange energies.
The observed relative changes of the correlation energies with increasing field strength
are also significantly smaller than those observed for the exchange energies.
One can therefore conclude that the importance of the correlation energy for the
binding properties of the 'weak-field ground state' ($1^10^+$) of helium decreases
with increasing field strength and that the exchange energy in case of the 'high-field
ground state' ($1^3(-1)^+$) of helium becomes increasingly important for the binding
properties whereas its correlation energy shows only some minor fluctuations.

We now turn to a discussion of the exchange and correlation energies for the ground states
of the hydrogen molecule in the presence of 
a magnetic field which is oriented parallel to the internuclear axis. 
Refs.\cite{Kra97,Det97,Det98} very recently clarified the question of the global ground 
states of the molecule for arbitrary field strengths. 
In field free space the ground state is well-bound and of $^1\Sigma_g^+$ symmetry, i.e.
in particular a spin singlet state. In the presence of a magnetic field
the spin-Zeeman shift lowers the total energy for triplet states with $M_s=-1$.
As a result a ground state crossover between the potential energy curves of the singlet
$1^1\Sigma_g$ and triplet $1^3\Sigma_u$ state occurs at $B \approx 0.18$.
For $B_{c_1} > 0.18$ the minimum of the potential energy curve of the 
$1^1\Sigma_g$ state lies higher in energy than the dissociation limit of 
the essentially repulsive (see below) potential energy curve of the $1^3\Sigma_u$ state.
The latter represents therefore the global ground state of the hydrogen molecule
if we increase the field strength beyond $B_{c_1}$.
The potential energy curve of the $1^3\Sigma_u$
state is a predominantly repulsive curve for any field strength.
Only for $B^{<}_{\sim} 1.0$ it exhibits a shallow outer van der Waals well
whose depth is of the order to $10^{-5}$ and contains no vibrational levels.
The hydrogen molecule is therefore unstable in the regime
of field strengths $B_{c_1} < B < B_{c_2}$ where $B_{c_2}=12.3$ is the field strength
for which a second crossover takes place.
The new ground state above $B_{c_2}$ is the state $1^3\Pi_u$ which is a strongly
bound state with a monotonically increasing binding energy with increasing field strength.

Due to antiparallel spins the 'weak-field ground state' $1^1\Sigma_g$ possesses
a vanishing exchange energy. The $1^3\Sigma_u$ state is also of little interest since
it exhibits only a very shallow outer well and does not correspond to a physically
bound state. Nevertheless we have provided in table 3 the positions of the outer
minimum as well as the exchange energies for the $1^3\Sigma_u$ state for field
strengths $B\le1.0$. The observed changes in the exchange energy are very minor
though the absolute values are pretty large. The latter fact has its origin 
in the exchange effects occuring because the inner part of the
wave function is very similar to the corresponding wave function
of the hydrogen molecular ion which is very different from the
dissociative wave function of separate atoms. 
Of much more interest is the 'high-field ground state' $1^3\Pi_u$ whose equilibrium
internuclear distances and exchange energies are given in table 3.
$R_{eq}$ decreases whereas the exchange energy increases by a factor of $20$
with increasing field strength $B=0-100$.
Figure 3 shows the quantity $R_{xb}$ for the $1^3\Pi_u$ state. It is almost
constant for $10^{-3}<B<10^{-1}$ and increases then approximately by a 
factor of two within the next two orders of magnitude of the field strength.
Thereafter, i.e. for $B^{>}_{\sim}10.0$, it shows only a minor decrease.
The origin of this observed increase of the exchange energy of the $1^3\Pi_u$ state
is very similar to the one observed for the 'high field ground states' of 
helium and lithium. In a HF-picture the $1^3\Pi_u$ state is described
by the united-atom orbitals $1s2p_{-1}$ which show a monotonically increasing binding
energy and decreasing size with increasing field strength. Consequently the
size of the molecule decreases also and its exchange energy increases due to the
increasing overlap of the orbitals (see above). We remark that showing instead
of the quantity $R_{xb}(B)$ the ratio $R_{xd}(B)$ of the exchange energy and the dissociation
energy for the $1^3\Pi_u$ state would have resulted first in a moderate increase 
(factor of $2.5$) with increasing field strength with a maximum at approximately $B \approx 2.0$ 
followed by a moderate decrease with further increasing strength of the field.

Finally we turn to a discussion of the correlation energies of the three states
involved in the ground states of the hydrogen molecule in a magnetic field.
For the 'weak-field ground state' $1^1\Sigma_g$ the correlation energy increases
by approximately a factor of three with increasing field strength $B=0-100$.
Figure 4 shows the corresponding ratio $R_{cb}(B)$ which is almost constant in
the regime $10^{-3}<B<10^{-1}$ and then decreases monotonically with increasing
field strength. The corresponding quantity $R_{cd}(B)$ which describes the
ratio of the correlation energy and the dissociation energy shows a very 
similar behaviour. The $1^3\Sigma_u$ state shows, according to table 4,
an extremely small contribution of the correlation energy which is almost of the
same size as our computational error. More interesting is again the
correlation energy of the $1^3\Pi_u$ state which decreases (see table 4) monotonically
by one order of magnitude in the complete range $B=0-100$. Apart from some
small amplitude oscillations the corresponding quantity $R_{cb}(B)$ is constant.
Again the ratio $R_{cd}(B)$ shows a similar behaviour, i.e. is also approximately
constant. 

\section{Summary and Conclusions}

We have investigated the behaviour of the exchange and correlation energies
for small atoms and molecules in magnetic fields covering the regime $B=0-100$.
To get the exchange energies Hartree-Fock calculations have been performed
for the helium and lithium atoms and for the hydrogen molecule using a flexible
mesh approach in order to adapt to the changing symmetries in the presence
of the external field. To extract the correlation energies we used a configuration
interaction approach which is based on an anisotropic generalized Gaussian basis
set particularly developed for the needs in a strong magnetic field.
Correlation energies have been determined for the helium atom and the hydrogen 
molecule. 

The overall tendency of the exchange energy with increasing field strength is
twofold. If less than two orbitals with parallel spins of the hydrogenic series $1s,2p_{-1},
3d_{-2},...$ are involved in the Hartree-Fock wave function then it can be 
expected that the exchange energy does not increase significantly with
increasing field strength and its relevance with respect to the binding
energy can be expected to decrease. An example for this situation is given
by the ground state of the lithium atom for weak magnetic fields, i.e.
the $1^20^+$ state which is constituted by the $1s^22s$ HF-configuration.
In ref.\cite{Iva98} it has been shown that, with increasing field strength,
the HF-configurations of the global ground states of an atom show
an increasing degree of spin polarization and, more important,
involve more and more of the orbitals belonging to the tightly bound hydrogenic
orbitals $1s,2p_{-1},3d_{-2},...$. The generic behaviour of the exchange energy
is then given by a monotonous, in many cases strong, increase with increasing
field strength. In fact the exchange energy increases for a typical case even faster than
the binding energy and becomes therefore increasingly relevant for the
binding properties of the atom. Examples for this behaviour are given
by the $1^3(-1)^+$ state of helium constituted by the $1s2p_{-1}$
HF-configuration and the $1^2(-1)^+$ and $1^4(-3)^+$ states of lithium which correspond to
the HF-configurations $1s^22p_{-1}$ and $1s2p_{-1}3d_{-2}$, respectively.
For the hydrogen molecule the strongly bound ground state in the 
high-field regime, i.e. the $1^3\Pi_u$ state, shows such a behaviour,
although it is not as well pronounced as in the case of the above-mentioned
states of helium and lithium.

So far we have summarized only the behaviour of the exchange energy for a given electronic
state. However, to get the complete picture one should take into account also
the fact that the ground state changes its symmetry with changing field strength.
The magnetic field increasingly spin-polarizes atoms and molecules with
increasing magnetic field strength, i.e. there occurs a series of ground state crossovers
which are due to spin flip transitions (note that these are not the only ground
state crossovers which actually occur but we have also crossovers without
spin flip transitions like for example the $1^20^+ \rightarrow 1^2(-1)^+$ crossover
in the case of the lithium atom). With each of those spin-flip transitions 
the number of unpaired electrons of the global ground state increases 
and gives a good chance to increase the corresponding exchange energy of the
global ground state in an abrupt way. This mechanism adds to the increasing
importance of the exchange energy for a given electronic state of a certain
symmetry with increasing field strength. In total one can therefore make the
general statement that with increasing field strength the exchange energy
becomes increasingly relevant for the binding properties of atoms and molecules.

For the correlation energy the situation is very different, indeed, it is in a sense almost the
opposite. Let us first discuss again the behaviour for an electronic ground state
of a given symmetry. The absolute value of the correlation energy increases in most cases 
with increasing field strength $B=0-100$ only by a minor factor $1-10$ which is much
less than the corresponding increase in exchange energy (c.f. the
ground states of helium or the hydrogen molecule).
Consequently, since in most cases the binding energy increases with increasing field strength,
the ratio of the correlation energy and the total binding energy
decreases or stays approximately constant. This means that the correlation energy becomes 
either increasingly irrelevant or at least does not become more relevant if we turn on
a stronger and stronger magnetic field. If we consider now the sequence of global
ground states with different symmetries which occurs with increasing field strength
then the absolute value of the correlation energy drops at each crossover which involves
more of the tightly bound orbitals of the hydrogenic series $1s,2p_{-1},3d_{-2},...$.
This feature adds to the decreasing importance of the correlation energy for
a given electronic ground state with a certain symmetry and increasing field strength.
One can therefore conclude that with increasing field strength the correlation energy
becomes increasingly irrelevant for the binding properties of atoms and molecules.

It is evident that there might be exceptions to the above-drawn picture which is
concluded from the few data available on small atoms and molecules. Further investigations
have to confirm and/or extend these conclusions. However this would take a considerable
effort in particular with respect to the development of exact methods for more than
two-electron systems which are currently not available.

\vspace*{0.5cm}

\begin{center}
{\bf{Acknowledgements}}
\end{center}
This work was supported by the National Science Foundation
through a grant (P.S.) for the Institute for Theoretical
Atomic and Molecular Physics at Harvard University and
Smithsonian Astrophysical Observatory.
M.V.I. and W.B. acknowledge financial support from the
Deutsche Forschungsgemeinschaft.

{}

\vspace*{1.0cm}

\begin{center}
{\bf Figure Captions}
\end{center}

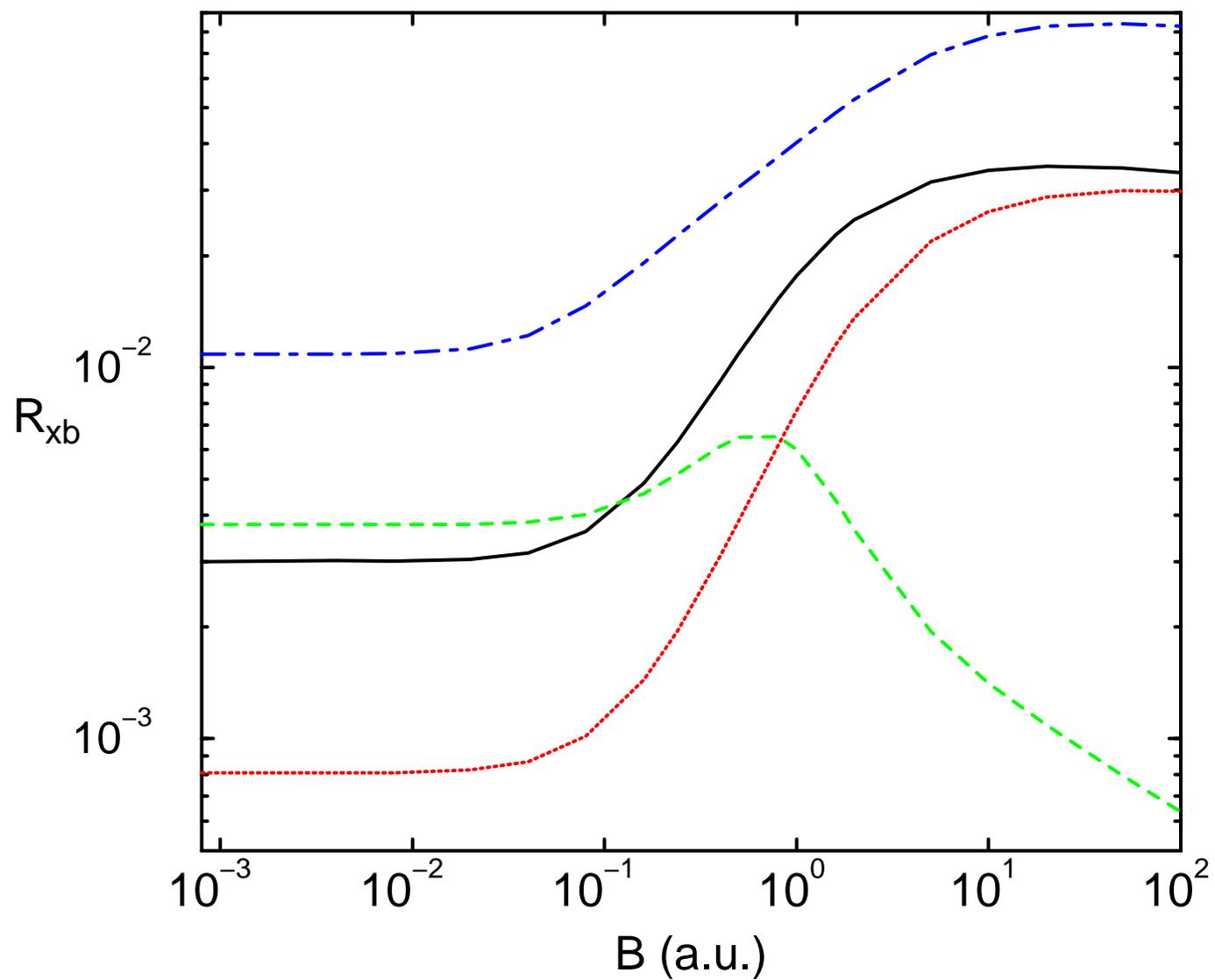
{\bf Figure 1:} The ratio $R_{xb}$ of the exchange energy and the total
binding energy as a function of the field strength $B$ on a double logarithmic scale:
for the high field ground state of helium $^3(-1)^+$
(solid line) and the three electronic states of lithium $^20^+$ (dashed line),
$^2(-1)^+$ (dotted line) and $^4(-3)^+$ (dash dotted line) which form the
ground states for arbitrary field strengths
($B = 1a.u.$ corresponds to $2.35 \times 10^{5} Tesla$). 

\vspace{0.5cm}

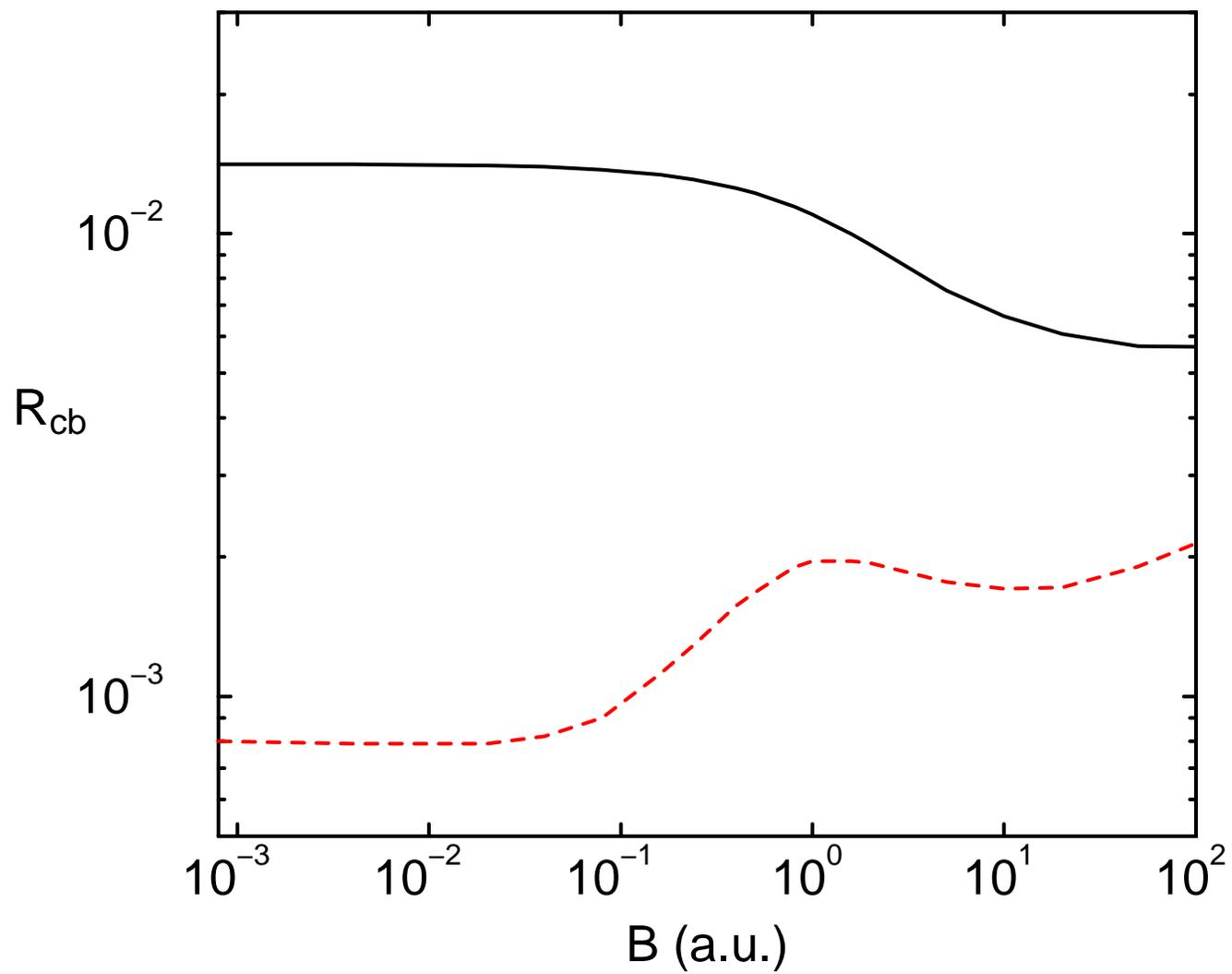
{\bf Figure 2:} The ratio $R_{cb}$ of the correlation energy and the total
binding energy as a function of the field strength $B$ on a double logarithmic scale:
for the two states $^10^+$ (solid line) and $^3(-1)^+$ (dashed line)
of helium forming the ground states for arbitrary field strengths.

\vspace{0.5cm}

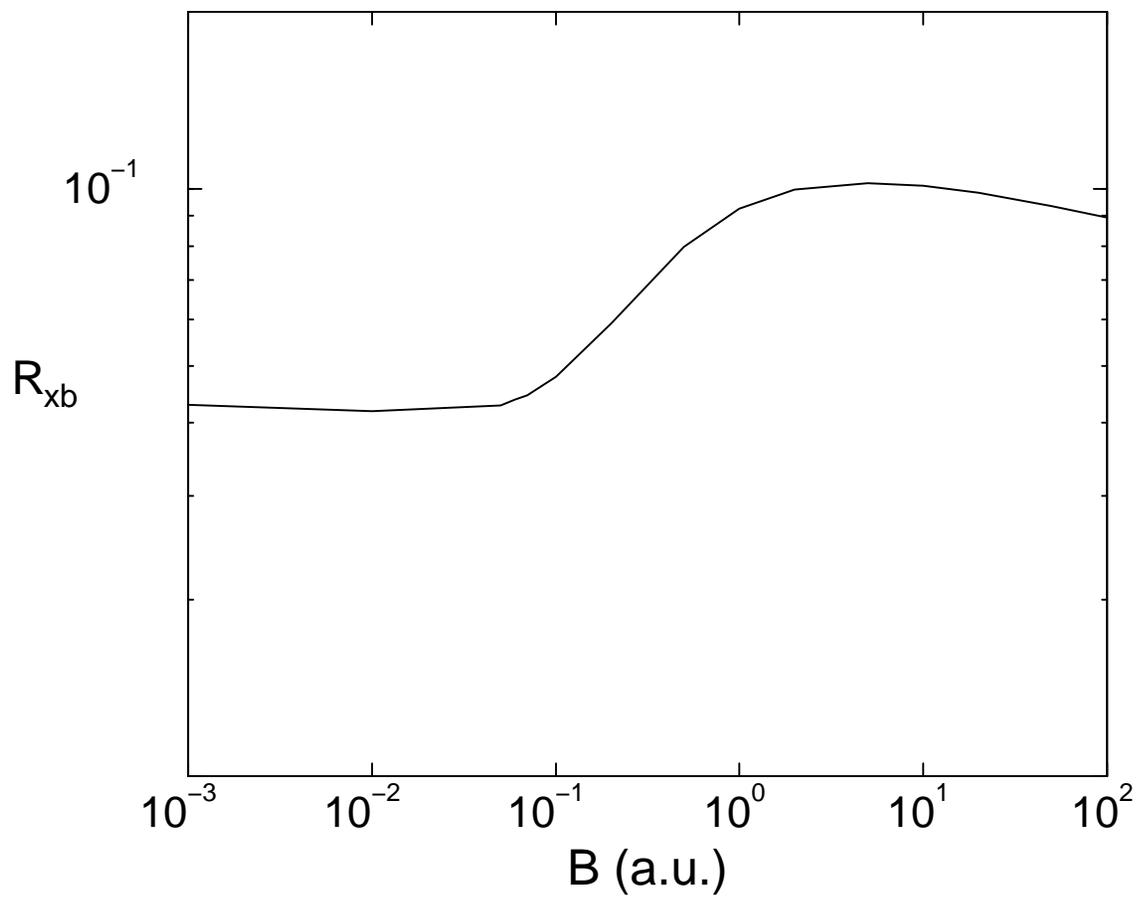
{\bf Figure 3:} The ratio $R_{xb}$ of the exchange energy and the total
binding energy as a function of the field strength $B$ on a double logarithmic scale
for the high field ground state $^3\Pi_u$ of the hydrogen molecule.

\vspace{0.5cm}

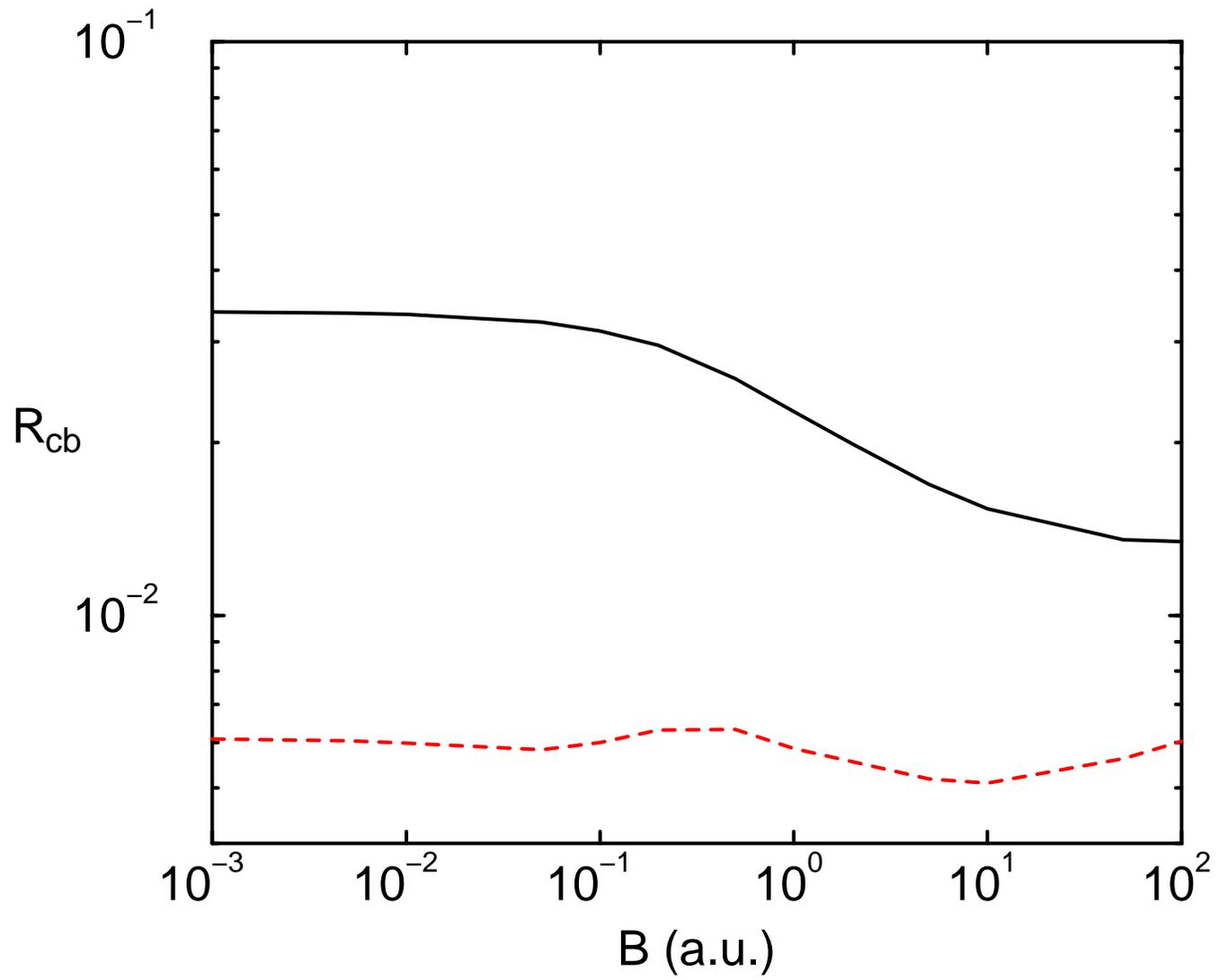
{\bf Figure 4:} The ratio $R_{cb}$ of the correlation energy and the total
binding energy as a function of the field strength $B$ on a double logarithmic scale:
for the electronic states $^1\Sigma_g$ (solid line) and $^3\Pi_u$
(dashed line) of the hydrogen molecule.

\vspace{1.0cm}

\begin{center}
{\bf Table Captions}
\end{center}

{\bf Table 1:} The exchange energies of the $^3(-1)^+$ state of helium and
the three electronic states $^20^+$, $^2(-1)^+$ and $^4(-3)^+$ of lithium
for field strengths in the range $B=0-100$.
All quantities are given in atomic units.

\vspace{0.5cm}

{\bf Table 2:} The correlation energies of the $^10^+$ and $^3(-1)^+$ states of helium 
for field strengths in the range $B=0-100$. 
All quantities are given in atomic units.

\vspace{0.5cm}

{\bf Table 3:} The exchange energies of the electronic $^3\Sigma_u$ and $^3\Pi_u$ states
of the hydrogen molecule at the corresponding equilibrium internuclear
distances $R_{eq}$ for field strengths in the range $B=0-100$.
All quantities are given in atomic units.

\vspace{0.5cm}

{\bf Table 4:} The correlation energies of the electronic
$^1\Sigma_g$, $^3\Sigma_u$ and $^3\Pi_u$ states
of the hydrogen molecule at the corresponding equilibrium internuclear
distances $R_{eq}$ for field strengths in the range $B=0-100$.
All quantities are given in atomic units.

\end{document}